\documentstyle[12pt,epsfig]{article}

\newskip\humongous \humongous=0pt plus 1000pt minus 1000pt
  \newif\ifdtup

\def\frac#1#2{ {{#1} \over {#2} }}

\def\beq{\begin{equation}}
\def\eeq{\end{equation}}

\def\non{\nonumber}
\def\beqn{\begin{eqnarray}}
\def\eeqn{\end{eqnarray}}

\def\L{\Lambda}

\def\as{\alpha_{\sf s}}
\def\a0{\alpha_{\sf 0}}
\def\aV{\alpha_{\sf V}}

\def\dm{\delta m}

\def\Tr{\mbox{Tr}\;}
\def\MSbar{\overline{\rm MS}}
\def\Dag#1{{#1}^\dagger}
\def\Real#1{\mbox{Re\,}({#1})}

\begin{document}
\begin{titlepage}
\begin{flushright}
     UPRF-2004-12 \\
     HU-EP-04/41\\
     August 2004\\
\end{flushright}
\par \vskip 10mm
\begin{center}
{\Large \bf
The $N_f=2$ residual mass \\
in Perturbative Lattice-HQET \\
for an improved determination of $m_b^{\MSbar}(m_b^{\MSbar})$}
\end{center}
\par \vskip 2mm
\begin{center}
F.\ Di Renzo$\,^a$,
and  L.\ Scorzato$\,^{b}$ \\
\vskip 5 mm
$^a\,${\it Dipartimento di Fisica, Universit\`a di Parma \\
and INFN, Gruppo Collegato di Parma, Italy}\\[.5 em]
\vskip 2 mm
$^b\,${\sl Institut f\"ur Physik, Humboldt Universit\"at, Berlin, Germany} \\[.5 em]
\end{center}
\par \vskip 2mm
\begin{center} {\large \bf Abstract} \end{center}
\begin{quote}
We determine to order $\alpha^3$ the so--called residual mass in the 
lattice regularisation of the Heavy Quark Effective Theory for $N_f=2$. 
Our (gauge--invariant) strategy makes use of Numerical Stochastic 
Perturbation Theory to compute the static interquark potential where 
the above mentioned mass term appears as an additive contribution. 
We discuss how the new coefficient we compute in the expansion of the 
residual mass can improve the determination of the ($\MSbar$) mass of the 
$b$--quark from lattice simulations of the Heavy Quark Effective Theory.
\end{quote}

\end{titlepage}

\section{Introduction}

Given our present computer resources, one can not accommodate the $b$--quark 
on the lattice keeping at the same time under control both finite size and 
finite lattice spacing effects. Despite this fact, the determination of 
the $b$--quark mass is one of the most precise results in Lattice Gauge Theory. 
There are various strategies that have been successfully applied to this 
purpose. A first determination comes from Non--Relativistic QCD 
\cite{NRQCD}. A second one relies on Heavy Quark Effective Theory (HQET) 
\cite{GGMR00}. Both these approaches have to deal with the mass counterterm 
whose computation we will be concerned with in this paper. In order to avoid 
this task - which has to be performed in Perturbation Theory (PT) - a third strategy 
has been devised, which makes use of HQET, but in a Non--Perturbative framework. 
A direct computation of the mass counterterm is avoided by making use of a 
finite scaling technique \cite{AlphaHQET}. A fourth method has been 
recently introduced which actually avoids HQET and remains within QCD. 
It again relies on a finite scaling technique, from which it takes the 
name of Step Scaling Method \cite{tvHQET}. 
A different approach - which can also profit of our computation - 
has been followed by \cite{BaliPineda}.
When different strategies are 
available, an important issue is the comparison of their results. In the 
quenched approximation, all the previous methods are consistent within errors. 
At the moment the only unquenched computation is the one in [2]. 
However the largest source of error, there, comes from the limited 
knowledge of  the perturbative residual mass. This can be 
strongly improved by the $\alpha^3$ computation which we are presenting in 
this paper. One important issue concerning the mass counterterm, whose 
perturbative computation we are going to present, is that it is a power-divergent 
quantity. Because of that, there is no continuum limit in PT, and that was the 
motivation for the approach in \cite{AlphaHQET}. Of course, while this is a 
conceptually important point, it is not a no-go for the first two methods. One 
has to stick to a finite value of the lattice spacing and to carefully assess 
the error which is inherent to this procedure. All these considerations set 
the stage for our computation. Our main task is to make it possible to extend 
the result in \cite{GGMR00} to a higher accuracy in the unquenched case. The 
quenched determination has already been greatly improved by a previous 
quenched computation of ours \cite{FrGg01} of which the present work is 
an extension. Having reached the goal of a higher accuracy in the analysis of 
\cite{GGMR00}, one can also investigate the status of PT with respect to the 
fact that one has to stick to a finite lattice spacing. \\
The structure of the paper is as follows. In the second section we 
sketch the theoretical framework, while in the third we present our 
computational strategy. More details about those issues in common with 
the quenched computation can be found in \cite{FrGg01}. The fourth section 
collects our results together with a brief discussion of the impact of 
the computation.

\section{The residual mass in Lattice HQET}

The Lagrangian density of HQET \cite{HQET} may be written as
\beq 
{\cal L}_{HQET} = \overline{h} D_4 h,
\eeq
where $D_4$ is the (gauge) covariant time derivative and $h$ is
the bi--spinor describing the only degrees of freedom of the heavy quark
that are relevant in this approximation. 
While the theory is appealing also in force of its simple form, it is well 
known that there are many subtleties one has to face, many of which have to 
do with a proper definition of the heavy quark mass itself. This is well 
evident in the fundamental relation one would like to exploit in order 
to connect the mass of a physical hadron ($M_B$) to the HQET expansion mass 
parameter ($m_b$) and the (linearly divergent) binding energy (${\cal E}$) 
\beq 
M_B =  m_b + {\cal E} + {\cal O}(1/m_b).  
\eeq
At this stage $m_b$ is actually not yet properly defined. Following 
\cite{GGMR00} (and references therein) one can proceed as follows. By 
matching the QCD propagator to its lattice HQET counterpart one gets a 
relation involving the pole mass 
\beq
m_b^{pole} = M_B - {\cal E} + \dm + {\cal O}(1/m_b).  
\eeq
In this relation a new character has entered the stage: a linearly divergent 
additive mass counterterm, the so--called residual mass $\dm$ which 
in lattice regularisation - contrary to dimensional regularisation -
is generated by quantum corrections.
Our task is now to compute this quantity in Perturbation Theory
\beq
\dm  =  \sum_{n \geq 0} \, \overline{X_n} \, \a0^{n+1}. 
\eeq
The pole mass (by writing $m_b^{pole}$ one has in mind the pole mass for the 
$b$--quark) can be related to the $\MSbar$ mass $\overline{m_b}=m_b^{\MSbar}$. The perturbative matching reads 
\beq \label{matching0}
\overline{m_b}(\overline{m_b}) = 
m_b^{pole} \, 
\left [ 1 + \sum_{n=0}^{\infty} (\frac{\as(\overline{m_b})}{\pi})^{n+1}
D_n \right ] \;.
\eeq
The coefficients $D_n$ with $n \leq 2$ are known \cite{CSR99,MR99}. One can 
now put everything together to get what the authors of \cite{GGMR00} call 
the master equation 
\beq \label{matching}
\overline{m_b}(\overline{m_b}) = 
\left [ M_B - {\cal E} + \sum_{n=0}^{\infty} (\as(\overline{m_b}))^{n+1}
\frac{X_n}{a}\right ]
\left [ 1 + \sum_{n=0}^{\infty} (\frac{\as(\overline{m_b})}{\pi})^{n+1}
D_n \right ] \;.
\eeq
After having arrived to this fundamental relation a few comments are in 
order. Note, first of all, that the last expression asks for both a 
non--perturbative (${\cal E}$) and a perturbative ($\dm$) computation. 
As it is well known, ${\cal E}$ can be computed as the decay constant 
of the correlation function of two axial currents. In Eq.~(\ref{matching}) 
many things are actually taking place, for which the perturbative expansion 
of the residual mass is responsible. First, $\dm$ is in charge of canceling the 
linear divergence of ${\cal E}$. Also, a subtle cancellation of 
renormalon ambiguities is taking place: the renormalon in the expansion of 
$\dm$ cancels the one in the perturbative relation between the pole and 
$\MSbar$ masses \cite{MS95}. It should now be stressed that both things are taking place 
in Perturbation Theory. Because of that, the cancellation of the linear 
divergence is incomplete and the continuum limit can not be taken. As for 
the renormalon ambiguities cancellation, the same coupling must be used 
in both perturbative expansions present in Eq.~(\ref{matching}). As we have already 
said, $D_2$ is known, which means that $\overline{X_2}$ (from which $X_2$ is computed) 
has to be computed in order to 
exploit at best the knowledge coming from continuum Perturbation Theory. 
$\overline{X_0}$ has been known for a long time, while $\overline{X_1}$ 
was computed in \cite{MS98}. 
As already said, $\overline{X_2}$ was computed in the quenched approximation in 
\cite{FrGg01} and then also in \cite{Trott}. Our goal is now to compute $\overline{X_2}$ 
for $N_f=2$ Wilson fermions, in order to make contact with the unquenched data of 
\cite{GGMR00}.

\section{The computational strategy}

As already said, the computational strategy is just the extension to $N_f=2$ 
of the computation in \cite{FrGg01}. 
Consider a gauge invariant loop such as a Wilson ($W$) or a Polyakov ($P$) 
loop. By following \cite{DV80}'s notation, one can write 
\beq
\langle W \rangle = \exp({- c \, L/a}) \; W_{log}. 
\eeq 
In the above formula $L$ is the length of the loop and $a$ the cutoff scale 
at small distance (of course one can think of the lattice spacing). $c$ is 
the linear divergence we are interested in. At the time of \cite{DV80} 
it was regarded as a mass renormalization of the test particle that one considers 
if one reads the loop as an effective amplitude.
The notation $W_{log}$ reminds that only logarithmic divergences are left. 
A first one is the divergence that can be absorbed in the definition of the 
renormalized coupling. A second logarithmic divergence appears if the contour 
has an angle; this is usually referred to as the corner divergence. Of course 
such a problem is there in the case of a Wilson loop and not for a Wilson line. 
One can now go to Lattice Perturbation Theory (LPT) and compute the static potential 
via Creutz's ratios
\beq \label{pot}
V(R) \equiv \lim_{T\rightarrow\infty} V_T(R) \;\;\;\;\;\;\;\;\;\;\;\;
V_T(R) \equiv \log \left( \frac{W(R,T-1)}{W(R,T)} \right).
\eeq
The corner divergences now disappear because of the ratio, the number of corners 
being the same in numerator and denominator. As for the coupling, 
there is a standard way to renormalize it:
\beq \label{VCoul}
V(R) = 2 \, \dm \, + \, V_{Coul}(R) \;\;\;\;\;\;
V_{Coul}(R) \equiv - C_F \frac{\aV(R)}{R}.
\eeq
This equation amounts to the definition of the coupling in the potential scheme. 
Since we are computing in LPT, what we are actually computing in the second formula 
of Eq.~(\ref{VCoul}) is the matching between the potential and the lattice 
couplings. This reads
\beq \label{VCoul2}
V(R) = 2 \, \dm \, - \frac{C_F}{R} \left( \a0 + c_1(R) \, \a0^2 + 
c_2(R) \, \a0^3 + \, \ldots \right)
\eeq
where the relevant coefficients are given by
\beqn
c_1(R) & = & 2 b_0 \log R + 2 b_0 \log \frac{\L_V}{\L_0}  \non \\ 
c_2(R) & = & {c_1(R)}^2 + 2 b_1 \log R + 2 b_1 \log \frac{\L_V}{\L_0} + 
\frac{b_2^{(V)}-b_2^{(0)}}{b_0}. \non 
\eeqn
The matching coefficients have been written down in terms of the dependence 
on $\L$ parameters and coefficients of the $\beta$-function. 
The good message is now that both the matching between the potential and 
the $\MSbar$ couplings (\cite{potms}) and the matching between the $\MSbar$ 
and the lattice couplings (\cite{lattms}) are known. This means that one 
can match the potential and the lattice couplings and as a result the only 
unknown quantity in Eq.~(\ref{VCoul2}) is $\dm$. This defines our strategy: 
we computed $V(R)$ in LPT up to the third order and fitted (order by order) 
$\dm$ to make contact with Eq.~(\ref{VCoul2}). 

\begin{figure}[htb] 
\begin{center}
\mbox{\epsfig{figure=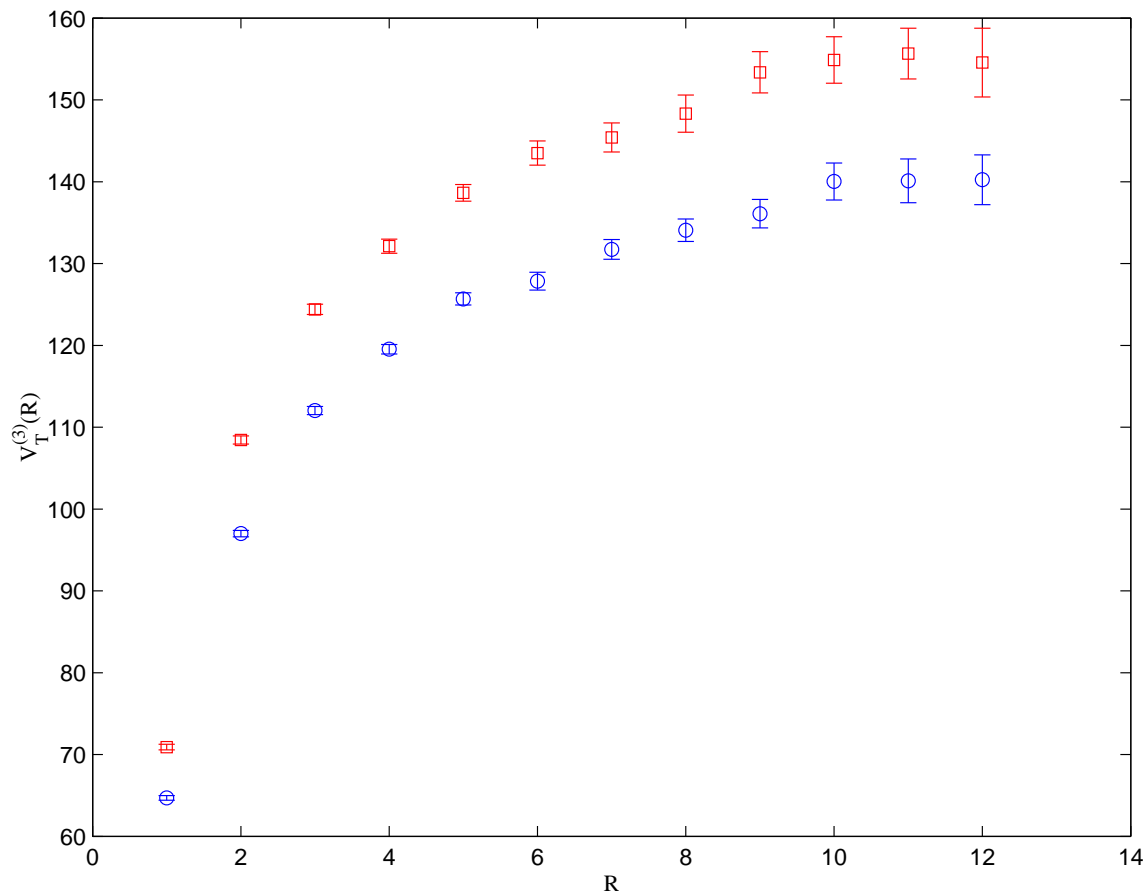,height=12cm}}
\caption
{Comparing $V_T^{(3)}(R)$ at $N_f=0$ (squares) and $N_f=2$ (circles). This is 
the third order of the potential with the expansion in terms of $\a0$.\label{fig:VvsV}}
\end{center}
\end{figure}

It is worth to comment that the only difference with respect to the procedure 
in \cite{FrGg01} comes from the dependence on $N_f$ of $c_1(R)$ and $c_2(R)$, 
which in turns comes from the dependence on $N_f$ of $\frac{\L_V}{\L_0}$ and 
$b_2^{(V)}-b_2^{(0)}$. The final effect of this can be seen\footnote{Both in this 
figure and in the following ones results are plotted in terms of an 
expansion in $\a0$. In \cite{FrGg01} the plots were in terms of $1/\beta$.} 
in Fig.~(\ref{fig:VvsV}). 

\subsection{Unquenching the computation}

Our computational tool was Numerical Stochastic Perturbation Theory 
(NSPT). This a stochastic way of performing Lattice Perturbation Theory 
which works by expanding the solution of the Langevin equation as a formal 
power series in the coupling
\[
U \rightarrow 1+ \sum_{k=1} \beta^{-k/2} U^{(k)}.
\]
Such a solution has to be plugged in the Langevin equation
\[
\frac{\partial}{\partial t} U \, = \, 
\left( -i \nabla S[U] -i \eta \right) U,
\]
$\eta$ being a gaussian noise and $\nabla$ being the Lie derivative on the group 
that acting on the action produces the classical equation of motion. 
In the limit of the stochastic time $t$ going to infinite, expectation values 
with respect to the gaussian ($\eta$) noise reconstruct the results of expectation 
values with respect to the path integral measure \cite{PW}. 
Needless to say, the only difference in computing Wilson loops 
in the quenched and the unquenched case are the background gauge configurations, 
which should be computed with a new weight in the functional integral, \emph{i.e.}
\[
e^{-S_G} \mapsto e^{-S_G} \det M = e^{-(S_G-Tr \ln M)},
\]
$S_G$ being the Wilson gauge action and $M$ being the lattice Dirac operator. 
The unquenched version of NSPT requires in the Langevin equation the evaluation of 
\[
\nabla S_G - \nabla \Tr \ln M = \nabla S_G - \Tr ( (\nabla M) M^{-1}).
\]
The second term of the former expression can now be evaluated by a new source of 
noise ($\langle \xi_i \xi_j \rangle_\xi = \delta_{ij}$)
\[
\langle \Real{\Dag{\xi_l}(\nabla M)_{ln}M^{(-1)}_{nk} \xi_k }\rangle_\xi,
\]
in which a sum is understood on repeated indices. This is an old strategy for 
unquenched simulations \cite{Bat}, which in NSPT gets greatly simplified by a 
simple observation. In our method everything should be expanded as a power series 
in the coupling, which means that also the matrix $M^{(-1)}$ has to be computed as 
the inverse of
\[
M = M^{(0)} + \sum_{k>0} \beta^{-k/2} M^{(k)}.
\]
Now the inverse of this expression reads
\[
M^{-1} = {M^{(0)}}^{-1} - \beta^{-1/2} {M^{(0)}}^{-1} M^{(1)} {M^{(0)}}^{-1} + \ldots
\]
The form of the first non trivial order is enough to explain where the good point is. 
The only inverse to be computed is the zeroth-order, which is diagonal in momentum 
space and independent of the gauge field. The unquenched version of NSPT was first discussed in \cite{Lat00}. The 
computation of Wilson loops on which this work relies is its first actual application. 
It is worth to inspect again Fig.~(\ref{fig:VvsV}). The $N_f=0$ data are obtained from a sample of 
$120$ configurations\footnote{Actually these configurations are split into two sets with different discretizations of the time step in the Langevin equation. This is needed in order to perform a continuum stochastic time extrapolation.}, 
while those at $N_f=2$ from a sample 
of $200$. As one can see from the errorbars, the statistical fluctuations 
are practically insensitive to the presence of fermion loops.
A reader interested in some extra details on the method can refer to \cite{NSPTnew}.
The computation of Wilson loops is actually only one 
out of a variety of computations for which the configurations we used 
were generated for (other goals have to do with full QCD, and in particular 
mainly with renormalization constants). 

\section{Results}

\subsection{$\dm$ to third order for $N_f=2$}

\begin{figure}[htb] 
\begin{center}
\mbox{\epsfig{figure=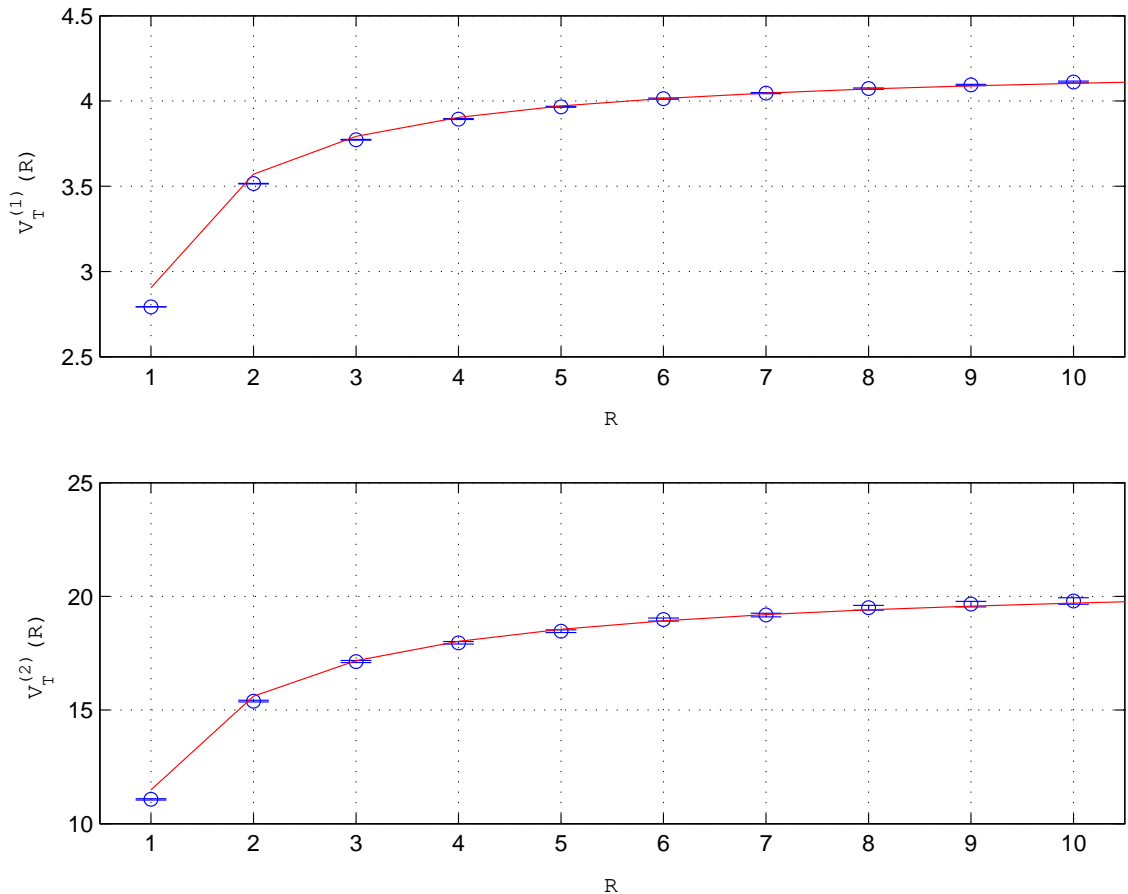,height=12cm}}
\caption
{A typical result of the fitting procedure described in the text. The 
plots refer to first and second order. 
The solid lines are 
the expressions of Eq.~(\ref{VCoul2}) once $\dm$ has been 
fitted, while circles are the $V_T(R)$. In this case $T=14$. 
Expansions are in terms of $\a0$.\label{fit_12}}
\end{center}
\end{figure}

\begin{figure}[htb] 
\begin{center}
\mbox{\epsfig{figure=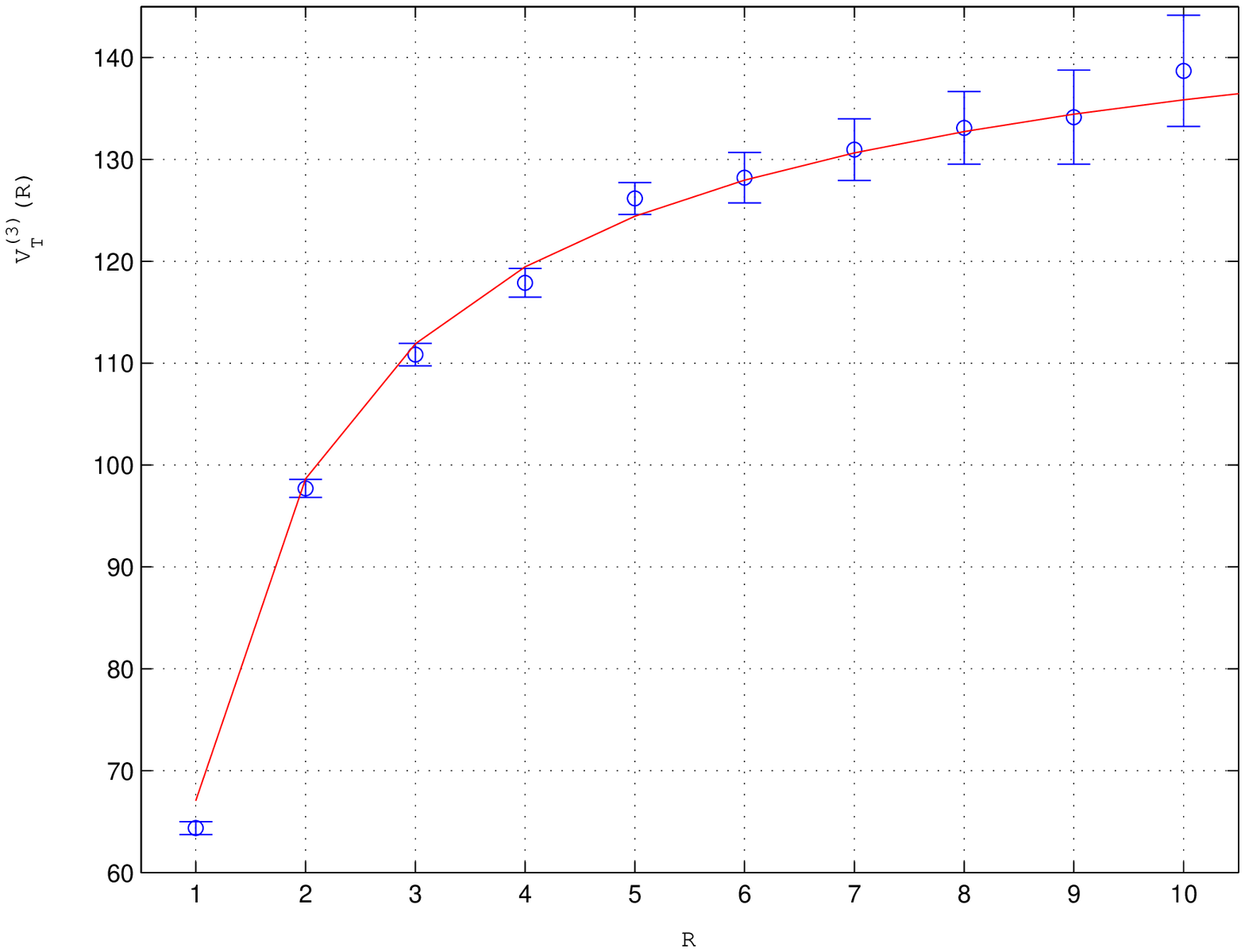,height=12cm}}
\caption
{The same as Fig.~(\ref{fit_12}) for the third order. Again, the expansion is in terms of $\a0$.\label{fit_3}}
\end{center}
\end{figure}

To perform the present computation we thermalized 200 configurations on a $32^4$ 
lattice with Wilson gauge action and $N_f=2$ Wilson fermions, which is the unquenching  that is relevant to supplement for an extension of the results in \cite{GGMR00}. 
100 configurations were produced at $dt=0.01$ (time step discretization of the Langevin stochastic time), 
and 100 at $dt=0.02$. The extrapolation errors at $dt=0$ are included in the errorbars in Fig.~(\ref{fig:VvsV}).  
The mass of both 
sea quark was set to zero, by exploiting the knowledge of $k_{crit}$, which 
- at this order - has been analytically computed in \cite{Haris,Pelo}. 
We computed $W(R,T)$ for all $R,T \leq 16$ up to order $\a0^3$. 
Again, a reader interested in the actual results for Wilson 
loops can refer to \cite{NSPTnew}. 
Before presenting our result, we now discuss in some more detail our 
procedure, together with some comments on errors. 
\begin{itemize}
\item
First of all, statistical errors are associated with the numerical nature 
of NSPT. These are magnified by the process of extracting the potential, 
which requires computing a series out of the direct expansions 
one obtains for the primary quantities (which are the Wilson loops). The 
statistical errors associated to our measures were computed via the 
bootstrap technique \cite{toyM}. 
\item
Given the finite lattice size nature of our computation, we can not actually 
take the $T\rightarrow\infty$ limit in Eq.~(\ref{pot}) and it is clear 
that Eq.~(\ref{VCoul2}) is distorted by lattice artifacts. Given the overall 
finite size and the presence of two scales (\emph{i.e.} $R$ and $T/R$) it 
follows that there is no simple finite scaling technique one can apply 
within the constraints of our computation. One issue is of course to take both 
$R$ and $T/R$ fairly large. One reassuring piece of information concerning lattice 
artifacts came from the fitting of the constants coefficients entering 
Eq.~(\ref{VCoul2}) (\emph{i.e.} $C_F$, $2 b_0 \log \frac{\L_V}{\L_0}$, \ldots). 
If one lets them as free parameters, their range of variation within the 
chosen value of $R$ and $T/R$ (see next point) is only of a few percent. 
\item
We determined the preferred values of the $\overline{X_i}$ by fitting our expansions of 
$V_T(R)$ to Eq.~(\ref{VCoul2}). In these fits the $\overline{X_i}$ were the only free parameters 
(\emph{i.e.} all the other constants were assigned their infinite volume, continuum 
limit values). The intervals in which fits were performed were such that three 
conditions were fulfilled: $R\geq3$, $T\geq12$ and $T > 2.5 \overline{R}$ (we denote by 
$\overline{R}$ the mean value of $R$ in the fitting interval; the fitting intervals 
themselves ranged from $3$ up to $7$ points). It was reassuring 
to find out that under these conditions good $\chi^2$ values were found. 
\item
As for the values of $\chi^2$ in the fitting intervals we referred to in the 
previous point, a first observation is that they are quite 
high for the fit of $\overline{X_0}$: the best values are of order $6,7$. This is not 
surprising, given our \emph{fixed lattice artifacts approach}: measurements 
of $V^{(1)}$ are simply precise enough to clearly detect lattice artifacts effects. 
For the coefficients $\overline{X_1}$ and $\overline{X_2}$ we quote an error that embraces the range of 
values for which $\chi^2$ does not exceed the value $2$ (again, in the same 
fitting intervals in which both $R$ and $T/R$ were fairly large).
\end{itemize}
Fig.~(\ref{fit_12}) and Fig.~(\ref{fit_3}) show an example of our fits at each order. 


The final results which we obtained are $\overline{X_0}=2.118(3)$ 
(to be compared with the analytical 
$\overline{X_0}=2.118...$), $\overline{X_1}^{(N_f=2)}=10.58(4)$ 
(to compare with \cite{MS98} $\overline{X_1}^{(N_f=2)}=10.588...$) and finally
\begin{equation}
	\overline{X_2}^{(N_f=2)} = 76.7(6).
\end{equation}

\subsection{The impact of the result}
The impact of the quenched result of \cite{FrGg01} has been known since 
the Lattice 2000 conference \cite{Vitt}. As for the unquenched case, the 
authors of \cite{GGMR00} have now 
repeated their analysis for the $b$--quark mass taking into account the new 
result for $X_2^{(N_f=2)}$. This results in \cite{Vice,Rak}
\[
	\overline{m_b}(\overline{m_b})^{(\mbox{unque})} = (4.21 \pm 0.03 \pm 0.05 \pm 0.04) \;\; \mbox{GeV}
\]
The last error is the one taking into account the indeterminations in the perturbative 
matching, on which our result has an effect. It would be twice as large with no knowledge of $X_2$ (with a central value for the mass of $4.26$; see \cite{Vitt}). 
The new analysis \cite{Vice} pins down  different values for the quenched and 
unquenched results, even if they are compatible 
within errors, so that further investigation is needed. As for the finite lattice 
spacing dependence (remember that in this approach one sticks to finite values 
of the lattice spacing), it is not dramatic and it gets decreased by including the new 
term in the matching. On the other hand, the control on renormalon ambiguities seems 
quite firm.

\section{Conclusions}
We computed the perturbative expansion of the unquenched ($N_f=2$) residual 
mass term in lattice Heavy Quark Effective Theory to order $\a0^3$. 
This allows to improve the determination of the $b$--quark mass from (unquenched) 
lattice simulations of the HQET. Errors appear to be under a fairly good 
control and the impact on the final result for the quark mass is quite important, 
even if further investigation is needed if one wants to carefully assess the 
difference between quenched and unquenched results for the $b$--quark mass.
 
\section*{Acknowledgments}
\par\noindent
We are very grateful to V. Gimenez which on behalf of all the authors of \cite{GGMR00} 
has shared with us the results of their analysis after our new result has been taken 
into account. 
Our interest in the subject was first triggered some years ago by G. Martinelli and 
C.T. Sachrajda, from both of which we learned much on the subject. 
F.D.R. acknowledges support from both Italian MURST 
under contract 2001021158 and from I.N.F.N. under {\sl i.s. MI11}. 
L.S. is supported by DFG through the Sonderforschungsbereich
'Computational Particle Physics' (SFB/TR 9).



\begin{thebibliography}{99}
\bibitem{NRQCD}
See for example S.~Collins, ``The mass of the b quark from lattice NRQCD,'' 
in {\em Quark Confinement and the Hadron Spectrum, World Scientific}, 325 (2002).
arXiv:hep-lat/0009040.
\bibitem{GGMR00}
V.~Gimenez, L.~Giusti, G.~Martinelli and F.~Rapuano,
JHEP {\bf 0003} (2000) 018
[arXiv:hep-lat/0002007].
\bibitem{AlphaHQET}
J.~Heitger and R.~Sommer  [ALPHA Collaboration],
JHEP {\bf 0402} (2004) 022
[arXiv:hep-lat/0310035].
\bibitem{tvHQET}
G.~M.~de Divitiis, M.~Guagnelli, F.~Palombi, R.~Petronzio and N.~Tantalo
Nucl.\ Phys.\ B {\bf 675}, 309 (2003)
[arXiv:hep-lat/0305018].
\bibitem{BaliPineda}
G.~S.~Bali and A.~Pineda,
Phys.\ Rev.\ D {\bf 69} (2004) 094001
[arXiv:hep-ph/0310130].
\bibitem{FrGg01}
F.~Di Renzo and L.~Scorzato,
JHEP {\bf 0102}, 020 (2001)
[arXiv:hep-lat/0012011].
\bibitem{HQET}
M.~Neubert,
Phys.\ Rept.\  {\bf 245}, 259 (1994)
[arXiv:hep-ph/9306320].
\bibitem{CSR99}
K.~G.~Chetyrkin and M.~Steinhauser,
Phys.\ Rev.\ Lett.\  {\bf 83}, 4001 (1999)
[arXiv:hep-ph/9907509].
K.~G.~Chetyrkin and A.~Retey,
Nucl.\ Phys.\ B {\bf 583}, 3 (2000)
[arXiv:hep-ph/9910332].
\bibitem{MR99}
K.~Melnikov and T.~v.~Ritbergen,
Phys.\ Lett.\ B {\bf 482}, 99 (2000)
[arXiv:hep-ph/9912391].
\bibitem{MS95}
G.~Martinelli and C.~T.~Sachrajda,
Phys.\ Lett.\ B {\bf 354} (1995) 423
[arXiv:hep-ph/9502352].
\bibitem{MS98}
G.~Martinelli and C.~T.~Sachrajda,
Nucl.\ Phys.\ B {\bf 559}, 429 (1999)
[arXiv:hep-lat/9812001].
\bibitem{Trott}
H.~D.~Trottier, N.~H.~Shakespeare, G.~P.~Lepage and P.~B.~Mackenzie,
Phys.\ Rev.\ D {\bf 65}, 094502 (2002)
[arXiv:hep-lat/0111028].
\bibitem{DV80}
V.~S.~Dotsenko and S.~N.~Vergeles,
Nucl.\ Phys.\ B {\bf 169}, 527 (1980).
\bibitem{potms}
Y.~Schr\"{o}der,
Phys.\ Lett.\ B {\bf 447}, 321 (1999)
[arXiv:hep-ph/9812205].
\bibitem{lattms}
C.~Christou, A.~Feo, H.~Panagopoulos and E.~Vicari,
Nucl.\ Phys.\ B {\bf 525}, 387 (1998)
[Erratum-ibid.\ B {\bf 608}, 479 (2001)]
[arXiv:hep-lat/9801007].
\bibitem{PW}
G.~Parisi and Y.~s.~Wu,
Sci.\ Sin.\  {\bf 24}, 483 (1981).
\bibitem{Bat}
G.~G.~Batrouni, G.~R.~Katz, A.~S.~Kronfeld, G.~P.~Lepage, B.~Svetitsky and K.~G.~Wilson,
Phys.\ Rev.\ D {\bf 32}, 2736 (1985).
\bibitem{Lat00}
F.~Di Renzo and L.~Scorzato,
Nucl.\ Phys.\ Proc.\ Suppl.\  {\bf 94}, 567 (2001)
[arXiv:hep-lat/0010064].
\bibitem{NSPTnew}
F.~Di Renzo and L.~Scorzato,
JHEP {\bf 0411} (2004) 036
[arXiv:hep-lat/0408015].
\bibitem{Haris}
E.~Follana and H.~Panagopoulos,
Phys.\ Rev.\ D {\bf 63} (2001) 017501
[arXiv:hep-lat/0006001].
\bibitem{Pelo}
S.~Caracciolo, A.~Pelissetto and A.~Rago,
Phys.\ Rev.\ D {\bf 64} (2001) 094506
[arXiv:hep-lat/0106013].
\bibitem{toyM}
R.~Alfieri, F.~Di Renzo, E.~Onofri and L.~Scorzato,
Nucl.\ Phys.\ B {\bf 578} (2000) 383
[arXiv:hep-lat/0002018].
\bibitem{Vitt}
V.~Lubicz,
Nucl.\ Phys.\ Proc.\ Suppl.\  {\bf 94}, 116 (2001)
[arXiv:hep-lat/0012003].
\bibitem{Vice}
V. Gim\'enez, private communication. 
\bibitem{Rak}
P.~E.~L.~Rakow,
arXiv:hep-lat/0411036.
\end{thebibliography}
\end{document}